\documentclass[traditabstract]{aa}
\usepackage{graphicx}
\usepackage{microtype}
\usepackage{txfonts}
\usepackage{multirow}
\usepackage{booktabs}
\usepackage{tabularx}
\usepackage{natbib}
\bibpunct{(}{)}{;}{a}{}{,}
\usepackage[colorlinks=true,linkcolor=blue,citecolor=blue, urlcolor=blue]{hyperref}
\usepackage{pgf}
\usepackage{aas_macros}
\usepackage{geometry}
\usepackage{layouts}
\usepackage{pdflscape}
\usepackage{bm}

\usepackage{amsmath}	
\usepackage{amssymb}	
\usepackage{textcomp}
\usepackage[para]{threeparttable}

\def\wgdlens{WGD$\,$2038$-$4008}
\def\tdist{D_{\Delta t}}
\def\Dd{D_{\rm d}}
\def\Dds{D_{\rm ds}}
\def\Ds{D_{\rm s}}
\def\zd{z_{\rm d}}
\def\zs{z_{\rm s}}
\def\kmsMpc{\rm km\ s^{-1}\,Mpc^{-1}}
\def\kext{\kappa_{\rm ext}}
\def\finalddt{1.68^{+0.40}_{-0.38}}
\def\finalH{65^{+23}_{-14}}
\def\LCDM{$\Lambda$CDM}

\begin{document}

\title{TDCOSMO. XVI. Measurement of the Hubble Constant from the Lensed Quasar WGD$\,$2038$-$4008}

\titlerunning{TDCOSMO. XVI. Measurement of $H_{0}$ from WGD$\,$2038$-$4008}
 
\author{Kenneth C.~Wong\inst{\ref{utokyo}}
    \and
    Fr\'ed\'eric Dux\inst{\ref{eso}, \ref{epfl}}
    \and
    Anowar J.~Shajib\inst{\ref{uchicago},\ref{kicp}}\thanks{NHFP Einstein Fellow}
    \and
    Sherry H.~Suyu\inst{\ref{tum},\ref{mpa},\ref{asiaa}}
    \and
	Martin Millon\inst{\ref{kipac}}
    \and
    Pritom Mozumdar\inst{\ref{ucd}}
    \and
	Patrick R.~Wells\inst{\ref{ucd}}
	\and
	Adriano Agnello\inst{\ref{hartree}}
	\and
	Simon Birrer\inst{\ref{sbu}}
	\and
	Elizabeth J.~Buckley-Geer\inst{\ref{fermi},\ref{uchicago}}
	\and
	Fr\'ed\'eric Courbin\inst{\ref{epfl},\ref{iccub},\ref{icrea}}
	\and
	Christopher D.~Fassnacht\inst{\ref{ucd}}
	\and
	Joshua Frieman\inst{\ref{uchicago},\ref{kicp},\ref{fermi}}
	\and
	Aymeric Galan\inst{\ref{tum},\ref{mpa}}
	\and
	Huan Lin\inst{\ref{fermi}}
	\and
    Philip J.~Marshall\inst{\ref{kipac},\ref{slac}}
	\and
	Jason Poh\inst{\ref{uchicago},\ref{kicp}}
	\and
	Stefan Schuldt\inst{\ref{unimi},\ref{inafmi}}
	\and
	Dominique Sluse\inst{\ref{starinst}}
	\and
	Tommaso Treu\inst{\ref{ucla}}
    }

   \institute{Research Center for the Early Universe, Graduate School of Science, The University of Tokyo, 7-3-1 Hongo, Bunkyo-ku, Tokyo 113-0033, Japan\label{utokyo}\\
    \email{kcwong19@gmail.com}
    \and
	European Southern Observatory, Alonso de Córdova 3107, Vitacura, Santiago, Chile\label{eso}
    \and
    Institute of Physics, Laboratory of Astrophysics, Ecole Polytechnique F\'ed\'erale de Lausanne (EPFL), Observatoire de Sauverny, 1290 Versoix, 
	Switzerland\label{epfl}
	\and
	Department of Astronomy \& Astrophysics, University of Chicago, Chicago, IL 60637, USA\label{uchicago}
    \and
    Kavli Institute for Cosmological Physics, University of Chicago, Chicago, IL 60637, USA\label{kicp}
	\and
	  Technical University of Munich, TUM School of Natural Sciences, Physics Department, James-Franck-Stra\ss{}e~1, 85748 Garching, Germany\label{tum}
	\and
	Max-Planck-Institut f{\"u}r Astrophysik, Karl-Schwarzschild-Str.~1, 85748 Garching, Germany\label{mpa}
    \and
	Academia Sinica Institute of Astronomy and Astrophysics (ASIAA), 11F of ASMAB, No.1, Section 4, Roosevelt Road, Taipei 106216, Taiwan\label{asiaa}
    \and
    Kavli Institute for Particle Astrophysics and Cosmology, Department of Physics, Stanford University, Stanford, CA, USA\label{kipac}
 	\and
	Department of Physics and Astronomy, University of California, Davis, CA, 95616, USA\label{ucd}
	\and
    STFC Hartree Centre, Sci-Tech Daresbury, Keckwick Lane, Daresbury, Warrington (UK) WA4 4AD \label{hartree}
    \and
    Department of Physics and Astronomy, Stony Brook University, Stony Brook, NY 11794, USA
    \label{sbu}
    \and
    Fermi National Accelerator Laboratory, P. O. Box 500, Batavia, IL 60510, USA\label{fermi}
    \and
    ICC-UB Institut de Ci\`encies del Cosmos, University of Barcelona, Mart\'i Franqu\`es, 1, E-08028 Barcelona, Spain\label{iccub}
    \and
    ICREA, Pg. Llu\'is Companys 23, Barcelona, E-08010, Spain\label{icrea}
    \and
    SLAC National Accelerator Laboratory, Menlo Park, CA, USA\label{slac}
    \and
    Dipartimento di Fisica, Universit\`a  degli Studi di Milano, via Celoria 16, I-20133 Milano, Italy\label{unimi}
    \and
    INAF - IASF Milano, via A. Corti 12, I-20133 Milano, Italy\label{inafmi}
    \and
    STAR Institute, Quartier Agora - All\'ee du six Aout, 19c B-400 Li\'ege, Belgium\label{starinst}
    \and
    Department of Physics and Astronomy, University of California, Los Angeles, 430 Portola Plaza, Los Angeles, CA 90095, USA\label{ucla}
    }

\authorrunning{K. C. Wong et al.} 

\date{Received XXX; accepted YYY}


\abstract
{Time-delay cosmography is a powerful technique to constrain cosmological parameters, particularly the Hubble constant ($H_{0}$).  The TDCOSMO collaboration is performing an ongoing analysis of lensed quasars to constrain cosmology using this method.  
In this work, we obtain constraints from the lensed quasar~\wgdlens~using new time-delay measurements and previous mass models by TDCOSMO.  This is the first TDCOSMO lens to incorporate multiple lens modeling codes and the full time-delay covariance matrix into the cosmological inference.  The models are fixed before the time delay is measured, and the analysis is performed blinded with respect to the cosmological parameters to prevent unconscious experimenter bias.  We obtain $\tdist = \finalddt$ Gpc using two families of mass models, a power-law describing the total mass distribution, and a composite model of baryons and dark matter, although the composite model is disfavored due to kinematics constraints.  In a flat \LCDM~cosmology, we constrain the Hubble constant to be $H_{0} = \finalH\, \kmsMpc$.  
The dominant source of uncertainty comes from the time delays, due to the low variability of the quasar. Future long-term monitoring, especially in the era of the {\it Vera C. Rubin} Observatory's Legacy Survey of Space and Time, could catch stronger quasar variability and further reduce the uncertainties.  This system will be incorporated into an upcoming hierarchical analysis of the entire TDCOSMO sample, and improved time delays and spatially-resolved stellar kinematics could strengthen the constraints from this system in the future.
}

\keywords{Cosmology: cosmological parameters --
        Cosmology: distance scale --
        Gravitational lensing: strong
        }

\maketitle


\section{Introduction} \label{sec:intro}
The Hubble constant ($H_{0}$) is a key cosmological parameter that represents the present-day expansion rate of the Universe.  Its value has important implications for the age, matter and energy content, and future of the Universe.  In recent years, a discrepancy has emerged among various methods of constraining $H_{0}$ \citep[see e.g.,][for a recent review]{verde+2023}.  In particular, the results from the {\it Planck} mission find $H_{0} = 67.4 \pm 0.5~\kmsMpc$ \citep{planck2020} in a flat \LCDM~cosmology based on observations of the cosmic microwave background (CMB).  Other methods anchored in the early Universe find similar results \citep[e.g.,][]{macaulay+2019,schoneberg+2022,brieden+2023,madhavacheril+2024}.  However, other measurements based in the local Universe tend to find a higher value \citep[e.g.,][]{freedman+2019,pesce+2020,anand+2022,palmese+2024}.  The most precise constraint, from the Supernovae, $H_{0}$, for the Equation of State of Dark Energy (SH0ES) collaboration, using type Ia supernovae calibrated by the Cepheid distance ladder, is $H_{0} = 73.0 \pm 1.0~\kmsMpc$ \citep{riess+2022}.  These conflicting measurements may point to new physics beyond flat \LCDM~if they cannot be explained by systematic errors.

Strong gravitational lensing of variable sources, such as quasars or supernovae, can be used to constrain cosmological parameters through a technique known as "time-delay cosmography." By measuring the time delay between multiple lensed images of the source, it is possible to constrain the "time-delay distance" to the lens system and thus $H_{0}$ \citep{refsdal1964, suyu+2010}.  Typically, background quasars lensed by galaxies have been used for time-delay cosmography due to their brightness and variability on short timescales, along with the fact that galaxy-scale lenses are also more abundant on the sky and have a shorter time delay relative to cluster-scale lenses.  Recent results using lensed supernovae \citep{kelly+2023,pascale+2024} have shown promise, but such objects are still quite rare in comparison \citep[see][and references therein]{suyu+2024}.

The TDCOSMO collaboration has been using time-delay cosmography to determine $H_{0}$ in a way that is independent of and complementary to other methods.  The latest TDCOSMO results from a joint analysis of seven lensed quasars constrain $H_{0}$ to $\sim2\%$ precision that is in agreement with late-Universe probes such as SH0ES, assuming that the lens galaxies are accurately parameterized by a power-law or stars+dark matter mass profile \citep{millon+2020,shajib+2020,wong+2020}.  Relaxing this assumption gives an $\sim8\%$ constraint \citep{birrer+2020}, which is statistically consistent with both {\it Planck} and SH0ES.  In order to use this technique to better constrain $H_{0}$, more lensed quasars with inferred time-delay distances are needed to increase the sample size and reduce the current uncertainty.

In this paper, we present a measurement of the time-delay distance to the lensed quasar~\wgdlens~based on existing lens models \citep{shajib+2022} and new time-delay measurements. To date, this is the only lens in the TDCOSMO sample that has been modeled by two independent teams using different modeling codes, which are combined into a final inference to account for systematic differences between the two.  Since a time delay had not yet been measured when the models were developed in \citet{shajib+2022}, they are effectively fixed prior to the measurement of the time delay, and we keep the cosmological parameter inference blinded until the analysis is completed to prevent any unconscious experimenter bias.  The results were unblinded on May 14, 2024 after all primary authors consented, and the results were not subsequently modified in any way.
The addition of this lens to the TDCOSMO sample will improve the overall constraint on cosmology.

This paper is organized as follows.  We provide an overview of time-delay cosmography in Sect.~\ref{sec:tdc}.  We describe the data we use in Sect.~\ref{sec:data} and our procedure for the cosmographic inference in Sect.~\ref{sec:inference}.  We present our results in Sect.~\ref{sec:results}.  We summarize our findings in Sect.~\ref{sec:summary}.  Throughout this paper, all magnitudes given are in the AB system.  All parameter constraints given are medians and 16th and 84th percentiles unless otherwise stated.

\section{Time-delay cosmography} \label{sec:tdc}
In this section, we summarize the principles behind time-delay cosmography, and refer the reader to, e.g., \citet{Treu+2022}, \citet{treu+2023}, and \citet{birrer+2024} for recent comprehensive reviews.  Light rays emitted from the background source experience a delay in arrival time at the observer because of the gravitational lensing effect.  This time delay is the result of two physical effects: the excess path length of the deflected ray relative to the path it would take in the absence of lensing, and the time dilation arising from the ray passing through the gravitational potential of the lens \citep{refsdal1964,shapiro1964}.  The excess time delay for a given lensed image (relative to an unlensed image) is 
\begin{equation} \label{eq:excess_td}
t(\vec{\theta}, \vec{\beta}) = \frac{\tdist}{c} \phi(\vec{\theta},\vec{\beta}),
\end{equation}
where $\vec{\theta}$ is the image position, $\vec{\beta}$ is the source position, $\phi$ is the "Fermat potential" at the image position, $c$ is the speed of light, and $\tdist$ is the time-delay distance.  The Fermat potential is defined as
\begin{equation} \label{eq:fermat}
\phi(\vec{\theta},\vec{\beta}) \equiv \frac{(\vec{\theta} - \vec{\beta})^{2}}{2} - \psi(\vec{\theta}),
\end{equation}
where $\psi$ is the deflection potential (a scaled projection of the gravitational potential) that is related to the (scaled) deflection angle $\vec{\alpha}$ by $\nabla \psi \equiv \vec{\alpha}$.  The time-delay distance is defined as
\begin{equation} \label{eq:ddt}
\tdist \equiv (1+\zd) \frac{\Dd \Ds}{\Dds},
\end{equation}
where $\zd$ is the lens redshift, and $\Dd$, $\Ds$, and $\Dds$ are the angular diameter distances from the observer to the lens, the observer to the source, and the lens to the source, respectively.  Each of the angular diameter distances in Equ~(\ref{eq:ddt}) is inversely proportional to $H_{0}$, therefore $\tdist \propto H_{0}^{-1}$.

In order to constrain $\tdist$ (and therefore, $H_{0}$) from a lensed quasar, one must measure the time delay between at least one pair of images, and constrain the Fermat potential at the position of the same images from an accurate mass model of the lens.  In practice, the time delay is measured by monitoring the lens over a period of time and comparing the light curves of the multiple images to look for common features corresponding to the same brightness fluctuation at the source.  The mass model of the lens is generally constrained from high-resolution imaging data, either from space telescopes or ground-based telescopes with adaptive optics \citep[e.g.,][]{chen+2019}.

A complicating factor is the mass-sheet degeneracy (MSD), a mathematical transform that can change the time delays by a multiplicative factor $\lambda$ (along with a rescaling in the unobservable source position), while leaving other observables unchanged \citep[e.g.,][]{Falco+1985, Schneider+2013}.  This is equivalent to having an infinite sheet of mass in the lens plane with constant surface density $\kappa \equiv 1-\lambda$ (in units of the critical density for lensing), although it can arise from both an internal transformation of the lens mass profile ($\lambda_{\mathrm{int}}$) and mass external to the lens $(\lambda_{\mathrm{ext}} \equiv 1-\kext)$.  This can, therefore, rescale the inferred $\tdist$ and $H_{0}$ by the same factor $\lambda$ if not accounted for.

The internal transformation, $\lambda_{\mathrm{int}}$, can be incorporated into a particular parameterization of the lens mass profile that is maximally conservative with respect to the MSD \citep{birrer+2020} and can be more tightly constrained with resolved kinematics of the lens galaxy \citep[e.g.,][]{shajib+2023,yildirim+2023}.  Throughout this analysis, we assume $\lambda_{\mathrm{int}} = 1$, similar to previous TDCOSMO analyses for individual systems \citep[e.g.,][]{suyu+2013,wong+2017,rusu+2020,shajib+2020}.  Instead, we consider two model families (see Sect.~\ref{subsec:fermat}), which allows us to see if there are hints of $\lambda_{\mathrm{int}} \neq 1$.  A more general result allowing for the maximal degeneracy with respect to the MSD will be incorporated into a future joint TDCOSMO analysis with all of the analyzed lenses (TDCOSMO collaboration, in preparation).

The external convergence, $\kext$, can be constrained through an analysis of the mass along the line-of-sight (LOS) to the lens.  
By characterizing the overdensity of the LOS (typically using cosmological simulations calibrated by relative galaxy number counts), it is possible to generate a distribution of $\kext$ values that can be factored into the final constraint \citep[e.g.,][]{Greene+2013,rusu+2017,wells+2023,wells+2024}.

\section{Data} \label{sec:data}
The quadruply-imaged lensed quasar~\wgdlens~was discovered in the Dark Energy Survey (DES) footprint by \citet{agnello+2018} through a combined search with the Wide-Field Infrared Survey Explorer (WISE) and the {\it Gaia} mission.  The lens redshift is $\zd=0.2283$, and the source redshift is $\zs=0.777$ \citep{buckleygeer+2020}.  The uncertainties on $\zd$ and $\zs$ are small enough to have a negligible impact on the results, so we assume they are fixed throughout the analysis.

\subsection{Fermat potential} \label{subsec:fermat}
\begin{figure}
\includegraphics[width=\columnwidth]{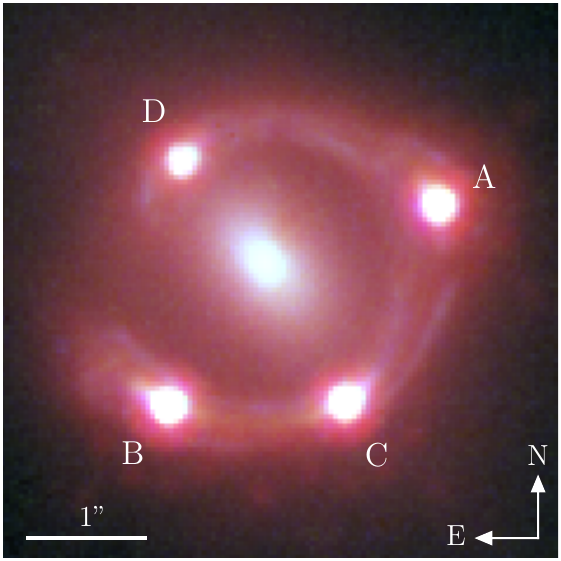}
\caption{\textit{HST} RGB color-composite of the system \wgdlens\ combining the imaging data in the F160W, F814W, and F475X filters. The four images A, B, C, and D are labeled. The scale bar represents 1$\arcsec$, and the arrows point to the north and east directions.}
\label{fig:wgd2038_img}
\end{figure}

\wgdlens~was modeled using imaging data from the Wide-Field Camera 3 (WFC3) on the {\it Hubble} Space Telescope (HST).  The data are taken in three filters: F475X and F814W in the ultraviolet-visual (UVIS) channel, and F160W in the infrared (IR) channel.  Fig.~\ref{fig:wgd2038_img} shows a composite color image of the lens system using these data \citep{shajib+2019}.

The lens was modeled by two independent teams in order to evaluate the consistency between two different lens modeling codes, {\sc glee} \citep{suyuhalkola2010,suyu+2012} and {\sc lenstronomy} \citep{birreramara2018,Birrer2021lenstro}.  
The details of the models are presented by \citet{shajib+2022}, and we summarize them here.  The two modeling teams each fit both an elliptical power-law total mass model and a stars$+$dark matter (hereafter, "composite") mass model, including external shear.  The models are constrained by the quasar image positions, as well as the surface brightness distribution of the lensed arcs from the quasar host galaxy in all three bands.  At the time of the analysis, a time delay had not yet been measured for~\wgdlens, so the results were presented in terms of the Fermat potentials and time delays predicted by the models at the quasar image positions.  We note that no modifications were made to the models from \citet{shajib+2022}, so the Fermat potentials are effectively fixed before the measurement of the time delay.

Despite working independently without knowledge of the other team's results, the power-law model results from both teams agreed to within $\sim1\sigma$.  
However, both teams independently found structural problems with their composite models (an extremely low dark matter fraction for the {\sc glee} team, and an unphysical velocity dispersion profile for the {\sc lenstronomy} team).  
The {\sc glee } team made the decision to combine the power-law and composite model results with equal weight, while the {\sc lenstronomy} team found that the predicted kinematics for their composite model were largely discrepant from the observed velocity dispersion, resulting in the power-law model receiving nearly all of the weight.  Despite these different choices, the final constraints on the Fermat potentials were in good agreement \citep[see][Fig. 21]{shajib+2022}.

\subsection{Stellar velocity dispersion and line of sight effects}
\label{ssec:ancillary}
A measurement of the velocity dispersion of the lens, as well as an analysis of the environment to constrain the external convergence ($\kext$) based on relative galaxy number counts and the external shear from the lens model, is presented in \citet{buckleygeer+2020}.  Both the kinematics and $\kext$ were incorporated into the lens models to determine the final Fermat potentials \citep{shajib+2022}.  As with the lens models, no modifications were made to the previously published results.  Only a measurement of the time delay of~\wgdlens~is then needed to constrain~$\tdist$~using time-delay cosmography (Eq.~\ref{eq:excess_td}).

The $\tdist$ posterior without $\kext$ and kinematics included will be used in an upcoming TDCOSMO hierarchical analysis (TDCOSMO collaboration, in preparation), as those observables will be incorporated for treatment at the population level.  For such future use cases, we present these results in Appendix~\ref{app:nokext_nokin}.

\subsection{Time delay} \label{subsec:td}

\wgdlens~was monitored in the $r$-band at two separate facilities, the Euler Swiss and ESO/MPG 2.2-m (2p2) telescopes, from April 2017 to October 2023. 
The general observing strategy adopted at the Euler telescope is detailed in~\citet{cosmograil2005} and~\citet{cosmograil05b}, and that of the 2p2 in ~\citet{cosmograil2018} and~\citet{cosmograil2019}.
Three seasons were accumulated at daily cadence on the 2p2, with the first in 2017, the second in 2021 shortened by the COVID-19 pandemic, and the last in 2022 temporarily interrupted by a snowstorm that forced the closing down of the La Silla observatory. 
Concurrently, the Euler telescope captured five more seasons, cadenced at a point every two days. 
The photometric deblending, estimated time delays, and uncertainties for~\wgdlens~and other lensed quasars will be presented in an upcoming TDCOSMO paper (Dux et al., in preparation), 
but we summarize the strategy and findings regarding the present target here.

\begin{figure*}
    \includegraphics[]{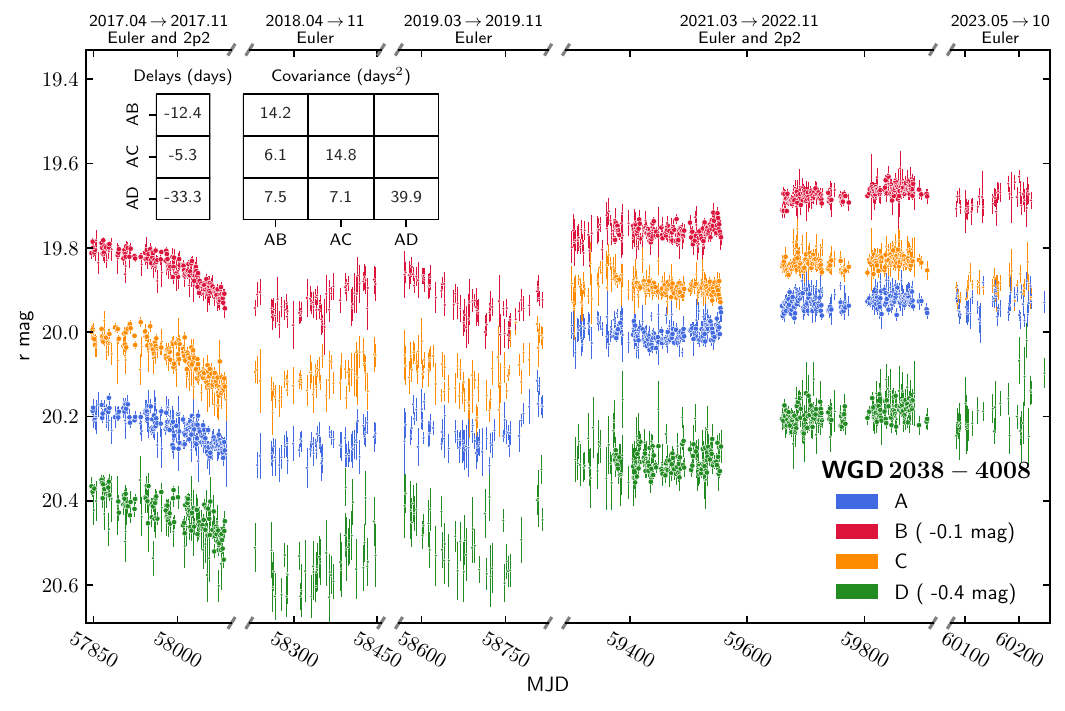}\vspace{-0.2cm}
    \caption{$r$-band light curves of \wgdlens, containing 3$\,$671 epochs of monitoring data. Data from the 2p2 and Euler telescopes are denoted by full disks and dots, respectively. When double coverage occurred, only the 2p2 data were used in the time-delay estimation.  The median delays relative to image A, along with the corresponding covariance matrix, are shown in the upper left corner.}
    \label{fig:light_curves}
\end{figure*}

Despite the combined 3$\,$671 epochs of monitoring data plotted in Fig.~\ref{fig:light_curves}, the~\wgdlens~quasar showed little variation, making  delay measurements challenging.
Nevertheless, comparing several initial guesses of the delay values from independent investigators showed an acceptable scatter, and their average was used as an initial starting point for value and uncertainty inference with the PyCS3 toolbox~\citep{pycs3}.
PyCS3 determines the values of the delays by optimizing the shifts between curves with different estimators and randomized initial shifts around the starting point.
For uncertainties, it creates mock light curves with multiple realizations of the observed noise power spectrum, also adding artificial modulations to the individual curves, mimicking microlensing.
The uncertainties are generally taken as the standard deviation of the resulting distributions of shifts. 
However, unlike previous $H_0$ estimations, where assuming uncorrelated errors in the delays was good enough to account for the uncertainty that maps into $H_0$, the present case required accounting for significant covariance among different delays.
Therefore, the delays were weighted by their full $3 \times 3$\footnote{
There are 6 distinct delay pairs in a quadruply lensed quasar, but because PyCS3 shifts all available curves together, there are only 3 degrees of freedom. 
Hence the $6\times6$ covariance matrix combining all delay pairs is singular, while any $3\times3$ sub-matrix containing all delays relative to one given lensed image contains the full information. 
}
covariance matrix in the delay likelihood (as opposed to a diagonal covariance matrix).
We provide the median time delays and covariance values with respect to image A, which we then use for the cosmographic inference (Sect.~\ref{sec:inference}). 
Our notation of the A$\,X$ delay, where $X$ can be B, C or D, means $\Delta t_{\rm AX} \equiv t_{\rm A}-t_{\rm X}$, where $t$ is given in Eq.~(\ref{eq:excess_td}). 
The resulting delay and covariance values are given at the top left of Fig.~\ref{fig:light_curves}.

We note that two possible time-delay solutions could fit the light curves equally well. In particular, the A$\,$C delay could either be negative ($-5.3$ days) or positive ($+7.9$ days), with only small repercussions on the values of the other delays. The change of sign between these two values is associated with a change in the order of arrival time between these two images. Both solutions were compared to the delays predicted by the mass models, and the solution with a negative A$\,$C delay was found to yield a much better $\chi^2$ with respect to the model-predicted time delays. 
Because the ordering of the arrival time predicted by the lens model is very robust, we discarded the positive A$\,$C solution and estimated the delay covariance matrix around the negative A$\,$C solution only.

Due to the faintness of the the lensed images (Vega $r$-magnitudes between 20 and 21) and the relative lack of variability of the source quasar during the monitoring period, the observational constraints on the delays are weak. Neglecting covariance, we find a precision of 30\%, 70\%, and 20\% on the A$\,$B, A$\,$C, and A$\,$D delays, respectively.

\section{Cosmographic inference} \label{sec:inference}
We combine the Fermat potentials with the new time-delay measurements in order to place constraints on $\tdist$.  As discussed in Sect.~\ref{subsec:td}, we use the delays relative to image A, along with the corresponding covariance matrix.  The $3\times3$ matrix contains all of the time-delay information, which we verify by testing the delays relative to the other images with their corresponding covariance matrices and finding the differences to be negligible.

The observational data $\mathcal{D}$ consists of the imaging data, kinematics, environment/LOS data, and time-delay measurements.  The likelihood of the data can be expressed as 
\begin{equation} \label{eq:lik}
\begin{aligned}
\mathcal{L}(\mathcal{D} \mid \tdist) = \int & \mathcal{L}(\mathcal{D}_{\mathrm{img,kin,LOS}} \mid \vec{\xi}_{\mathrm{model}},\kext)~\times \\
& \mathcal{L}(\mathcal{D}_{\mathrm{td}} \mid \vec{\xi}_{\mathrm{model}},\kext,\tdist)~\times \\
& p(\vec{\xi}_{\mathrm{model}},\kext)\ \mathrm{d}\vec{\xi}_{\mathrm{model}}\ \mathrm{d}\kext.
\end{aligned}
\end{equation}
Here, $\vec{\xi}_{\mathrm{model}}$ are the parameters of the lens model, including the lens and source light distributions. The probability density function $p(\vec{\xi}_{\mathrm{model}},\kext)$ is the prior on the model and external convergence parameters.  In practice, the sampling of the posterior can be split into separate parts (e.g., the lens model, kinematics, LOS).  All of the samplings aside from the time-delay likelihood were performed in \citet{shajib+2022}, and we include the time-delay likelihood as the final piece of the inference in this work.  We sample $\tdist$ assuming a uniform prior in the range $0 \leq \tdist \leq 4$ Gpc.

In addition to the constraint on $\tdist$, which is independent of cosmology, we also constrain $H_{0}$ directly in a flat \LCDM~cosmology.  We assume a uniform prior in the range $0 \leq H_{0} \leq 150~\kmsMpc$.  We fix $\Omega_{\mathrm{m}} = 0.3$ and $\Omega_{\Lambda} = 0.7$, as they are very weakly constrained by time-delay cosmography applied to a single system \citep[e.g.,][]{bonvin+2017,wong+2020}.

We first perform the inference for the {\sc glee} and {\sc lenstronomy} models separately to evaluate the variation in the result due to the differences between lens modeling codes.  For our final result, we need to combine the model constraints from {\sc glee} and {\sc lenstronomy}.  Statistically weighting the models from the two codes is complicated by the different choices made by each modeling team, including differences in the image reconstruction region and differences in the PSF reconstruction, among other things.  Instead, we take the conservative approach of combining the model constraints from {\sc glee} and {\sc lenstronomy} with equal weight to marginalize over any systematic differences between the two codes.  This is the first TDCOSMO lens to incorporate multiple lens models into the final inference in this way.

The time-delay distance and the corresponding $H_{0}$ value are kept blinded until the time-delay measurement is finalized and all primary authors agree to unblind. This is done to prevent unconscious experimenter bias.  We do this by either subtracting the median of the distributions or setting the median value to zero in any plots and tables viewed during the analysis.  In this way, we are still able to view the spread of the distribution of these parameters and their covariances with other parameters, but not their absolute values until the analysis is finished.  Once the results were unblinded on May 14, 2024, no further modifications were made to any part of the analysis.

In principle, it is possible to constrain the angular diameter distance to the lens, $\Dd$, by combining the lens model constraints with kinematics of the lens galaxy \citep[e.g.,][]{paraficzhjorth2009,jee+2015,jee+2016}, as has been done for some of the lenses in the TDCOSMO sample \citep{birrer+2019,chen+2019,jee+2019}.  Since the models used in this work were already completed in \citet{shajib+2022}, and incorporating $\Dd$ would require a new inference that would change the model weighting, we do not include $\Dd$ in order to maintain our blind analysis.  The constraints from $\Dd$ tend to be weaker than those from $\tdist$ when based on single aperture-averaged kinematic measurements of the lens galaxies \citep[see e.g.,][Table 2]{wong+2020}, so they are unlikely to have a large impact on our results.

\section{Results} \label{sec:results}
In this section, we present the results of combining the new time delays with the Fermat potentials from \citet{shajib+2022}, including the $\kext$ and lens velocity dispersion from \citet{buckleygeer+2020}.  As previously mentioned, the results without $\kext$ and kinematics included are provided in Appendix~\ref{app:nokext_nokin}.

\subsection{Time-delay distance} \label{subsec:ddt_result}
\begin{figure}
\includegraphics[width=\columnwidth]{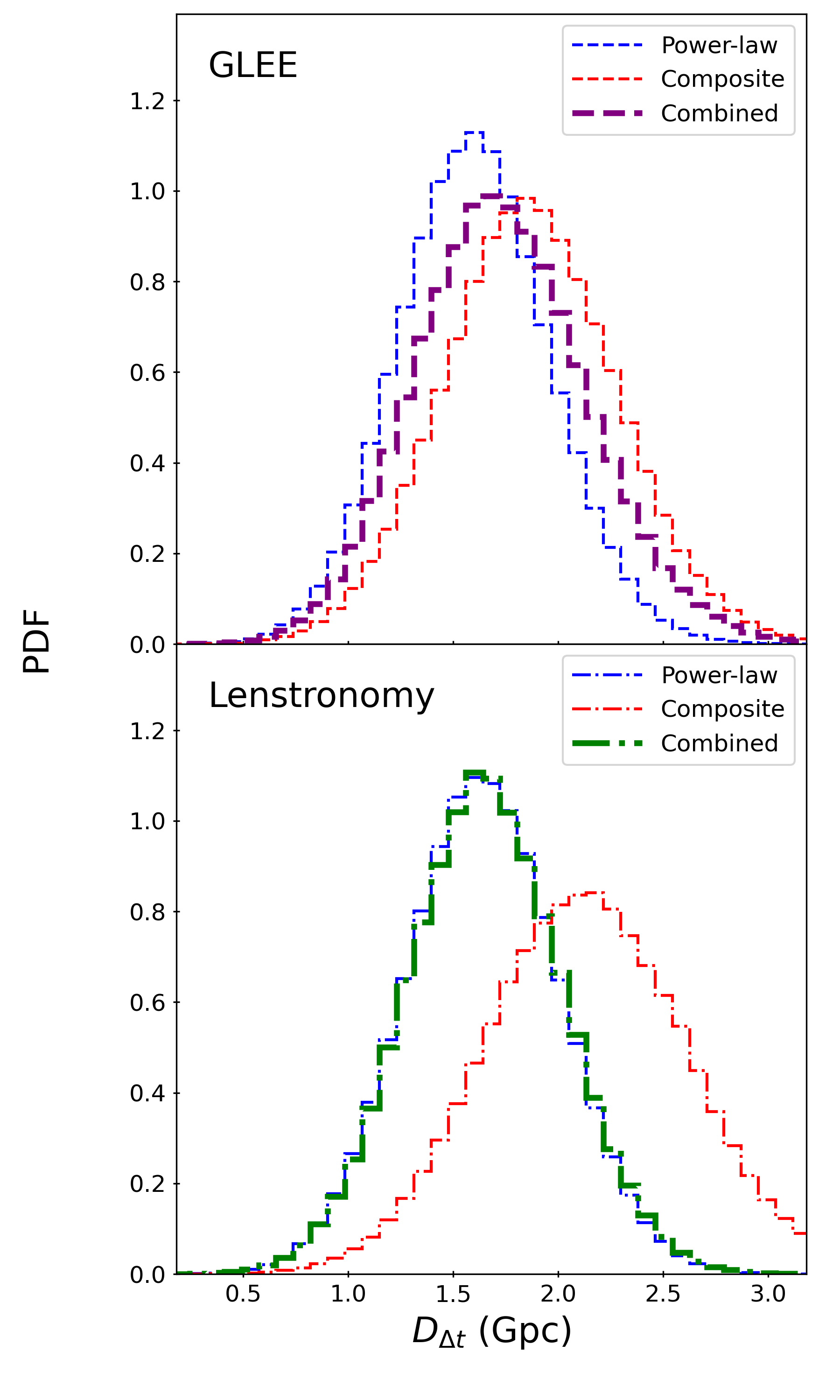}
\caption{$\tdist$ distribution for {\sc glee} (top) and {\sc lenstronomy} (bottom) for different mass models.  The power-law (blue) and composite (red) model constraints are shown separately, as well as the combined constraint (purple and green, respectively) for each code.  The {\sc glee} results give equal weight to the power-law and composite models, while the {\sc lenstronomy} results give most of the weight to the power-law model based on the kinematics weighting.  A flat prior in the range $0 \leq \tdist \leq 4$ Gpc is assumed.}
\label{fig:ddt_comparison}
\end{figure}

Using a uniform $\tdist$ prior (Sect.~\ref{sec:inference}), we first calculate the $\tdist$ distributions for {\sc glee} and {\sc lenstronomy} separately.  In Fig.~\ref{fig:ddt_comparison}, we show the power-law model, composite model, and combined results for each of the two modeling codes.  As described in Sect.~\ref{subsec:fermat} and \citet{shajib+2022}, the combined results for {\sc glee} give equal weight to the power-law and composite models, while the combined results for {\sc lenstronomy} give most of the weight to the power-law model based on kinematic constraints.  The results for both modeling codes are in good agreement, and are presented in detail in Table~\ref{tab:ddt}.

\renewcommand*\arraystretch{1.5}
\begin{table}
\caption{Time-delay distance constraints.\label{tab:ddt}}
\begin{minipage}{\linewidth}
\begin{tabular}{l|c}
\hline
\multirow{2}{*}{Model} &
$D_{\Delta t}$
\\
&
(Gpc)
\\
\hline
{\sc glee} power-law &
$1.60_{-0.35}^{+0.36}$\\
{\sc glee} composite &
$1.86_{-0.40}^{+0.41}$\\
{\sc glee} combined &
$1.72_{-0.39}^{+0.42}$\\
\hline
{\sc lenstronomy} power-law &
$1.64_{-0.36}^{+0.37}$\\
{\sc lenstronomy} composite &
$2.13_{-0.47}^{+0.48}$\\
{\sc lenstronomy} combined &
$1.65_{-0.36}^{+0.37}$\\
\hline
{\sc glee}+{\sc lenstronomy} final &
$1.68_{-0.38}^{+0.40}$\\
\hline
\end{tabular}
\\
\tablefoot{Reported values are medians, with errors corresponding to the 16th and 84th percentiles.  A flat prior in the range $0 \leq \tdist \leq 4~\mathrm{Gpc}$ is assumed.}
\end{minipage}
\end{table}
\renewcommand*\arraystretch{1.0}

\begin{figure}
\includegraphics[width=\columnwidth]{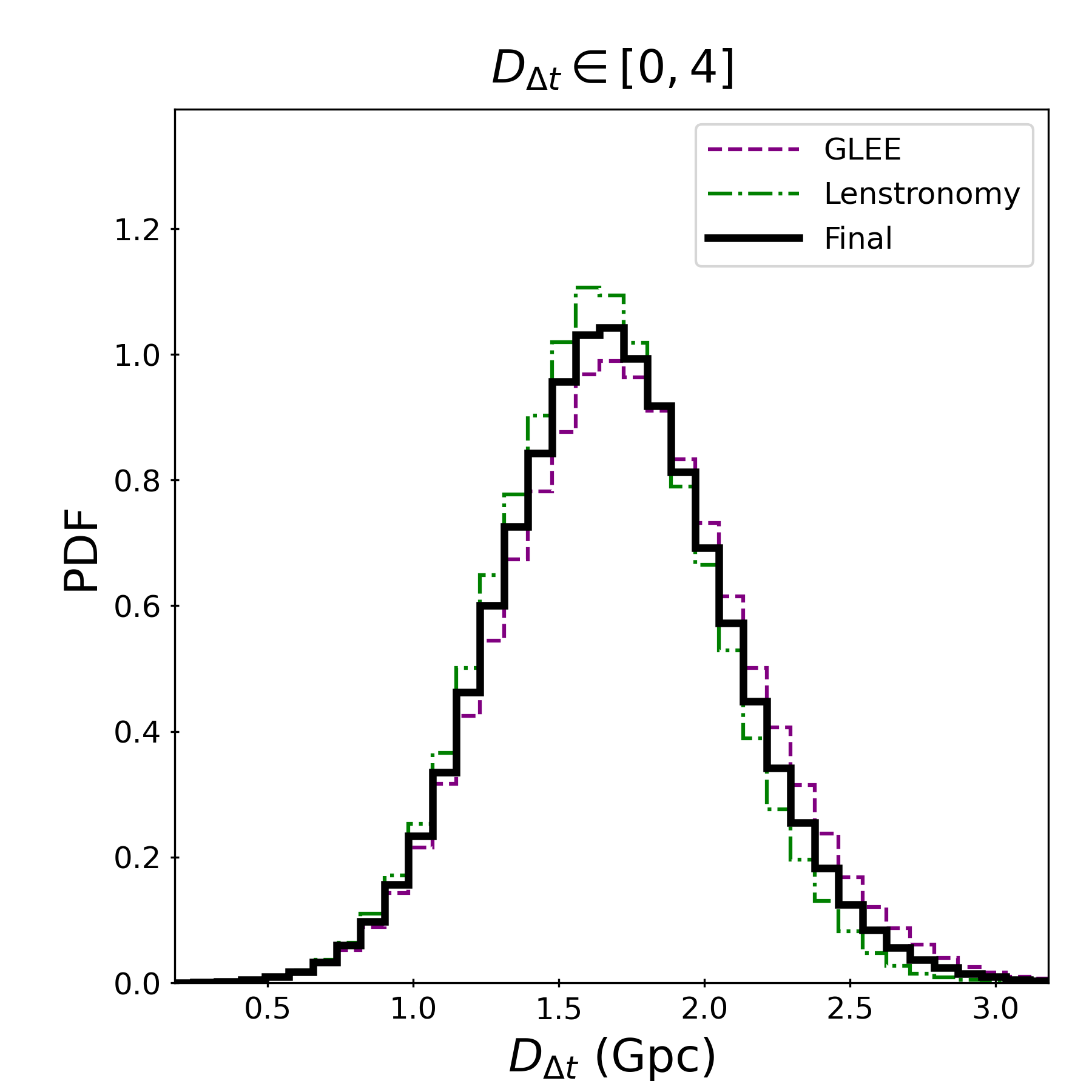}
\caption{$\tdist$ distribution using a flat prior in the range $0 \leq \tdist \leq 4$ Gpc.  Shown are the results for {\sc glee} (purple), {\sc lenstronomy} (green), and the final combined result (black).}
\label{fig:ddt_combined}
\end{figure}

Figure~\ref{fig:ddt_combined} shows the final joint constraint on $\tdist$, weighting the combined {\sc glee} and {\sc lenstronomy} models equally.  We find $\tdist = \finalddt$ Gpc.  This result is also presented in Table~\ref{tab:ddt}.  The large uncertainty on this result is dominated by the time-delay measurement uncertainty, which ranges from $20 - 70\%$ for the different image pairs resulting from the low variability of the source quasar.  By comparison, the uncertainty on the Fermat potential differences between image pairs, including LOS and kinematics, ranges from just $7 - 13\%$.

\subsection{$H_{0}$ in flat~\LCDM} \label{subsec:h0_result}
\begin{figure}
\includegraphics[width=\columnwidth]{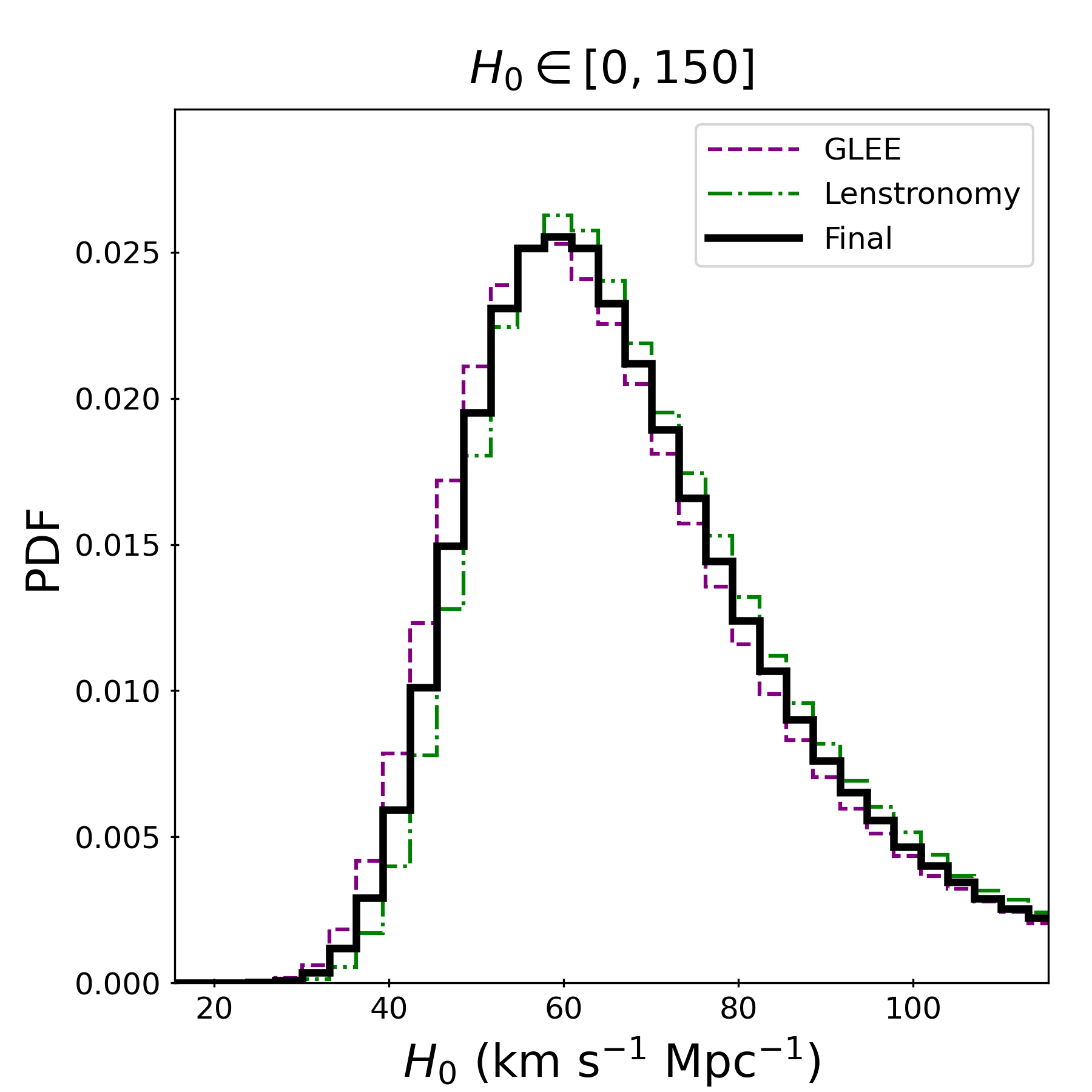}
\caption{$H_{0}$ distribution for flat~\LCDM~using a flat prior in the range $0 \leq H_{0} \leq 150~\kmsMpc$.  We assume $\Omega_{\mathrm{m}} = 0.3$ and $\Omega_{\mathrm{\Lambda}} = 0.7$.  Shown are the results for {\sc glee} (purple), {\sc lenstronomy} (green), and the final combined result (black).}
\label{fig:h0_combined}
\end{figure}

Assuming a uniform prior on $H_{0}$ with fixed $\Omega_{\mathrm{m}} = 0.3$ and $\Omega_{\mathrm{\Lambda}} = 0.7$, we find $H_{0} = \finalH\, \kmsMpc$ in a flat~\LCDM~cosmology.  The final distribution, along with the individual {\sc glee} and {\sc lenstronomy} constraints, is shown in Fig.~\ref{fig:h0_combined}, and the statistics for all distributions are presented in Table~\ref{tab:h0}.  As with the $\tdist$ constraint, the uncertainty comes primarily from the time-delay measurement.

\renewcommand*\arraystretch{1.5}
\begin{table}
\caption{$H_{0}$ and $\tdist$ constraints in a flat \LCDM~cosmology.\label{tab:h0}}
\begin{minipage}{\linewidth}
\begin{tabular}{l|cc}
\hline
\multirow{2}{*}{Model} &
$H_{0}$ &
$D_{\Delta t}$
\\
&
($\mathrm{km~s^{-1}~Mpc^{-1}}$) &
(Gpc)
\\
\hline
{\sc glee} power-law &
$69_{-14}^{+23}$ &
$1.43_{-0.36}^{+0.36}$\\
{\sc glee} composite &
$59_{-12}^{+21}$ &
$1.66_{-0.43}^{+0.43}$\\
{\sc glee} combined &
$64_{-14}^{+22}$ &
$1.53_{-0.40}^{+0.43}$\\
\hline
{\sc lenstronomy} power-law &
$67_{-14}^{+23}$ &
$1.46_{-0.37}^{+0.38}$\\
{\sc lenstronomy} composite &
$52_{-11}^{+18}$ &
$1.88_{-0.49}^{+0.50}$\\
{\sc lenstronomy} combined &
$67_{-14}^{+23}$ &
$1.47_{-0.38}^{+0.38}$\\
\hline
{\sc glee}+{\sc lenstronomy} final &
$65_{-14}^{+23}$ &
$1.50_{-0.39}^{+0.40}$\\
\hline
\end{tabular}
\\
\tablefoot{Reported values are medians, with errors corresponding to the 16th and 84th percentiles.  A flat prior in the range $0 \leq H_{0} \leq 150~\mathrm{km~s^{-1}~Mpc^{-1}}$ is assumed.  We fix $\Omega_{\mathrm{m}} = 0.3$ and $\Omega_{\Lambda} = 0.7$.}
\end{minipage}
\end{table}
\renewcommand*\arraystretch{1.0}

\section{Summary} \label{sec:summary}
We present new time-delay measurements for the lensed quasar~\wgdlens.  These delays are combined with the Fermat potentials from previously developed lens models to constrain the time-delay distance and Hubble constant in flat \LCDM.  The result combines models that assume either a power-law mass profile or a composite stars+dark matter profile, although the composite models were either down-weighted due to kinematics or showed unusual behavior in our previous analysis \citep{shajib+2022}.  The analysis is performed blindly with respect to $\tdist$ and $H_{0}$.  This is the first lens in the TDCOSMO sample to incorporate multiple lens models into the final result to account for systematic differences between lens modeling codes.  It is also the first TDCOSMO lens to use the full time-delay covariance matrix in the cosmographic inference.

We determine the time-delay distance of~\wgdlens~to be $\tdist = \finalddt$ Gpc.  We constrain the Hubble constant to be $H_{0} = \finalH\, \kmsMpc$ in flat~\LCDM, assuming $\Omega_{\mathrm{m}} = 0.3$ and $\Omega_{\Lambda} = 0.7$.  The dominant source of uncertainty in the cosmographic inference from this system comes from the time-delay measurements due to the low variability of the quasar.  Future long-term monitoring, especially in the era of {\it Vera C. Rubin} Observatory's Legacy Survey of Space and Time, could reduce the uncertainties by catching stronger quasar variability.  Spatially-resolved kinematics of the lens galaxy could also improve the mass model constraints, particularly for the composite models.

Although the constraints from~\wgdlens~alone are too broad to shed light on the $H_{0}$ tension, it will be included as part of the full TDCOSMO sample in a future hierarchical analysis that will provide the most robust constraint on $H_{0}$ (in flat \LCDM~as well as other cosmologies) from time-delay cosmography to date (TDCOSMO collaboration, in preparation).

\begin{acknowledgements}
This work is supported by JSPS KAKENHI Grant Numbers JP20K14511, JP24K07089.
This work received support from the Swiss National Science Foundation (SNSF) and by the European Research Council (ERC) under the European Unions Horizon 2020 research and innovation programme (COSMICLENS: grant agreement No 787886). 
This work was supported by NASA through the NASA Hubble Fellowship grant HST-HF2-51492 awarded to AJS by the Space Telescope Science Institute, which is operated by the Association of Universities for Research in Astronomy, Inc., for NASA, under contract NAS5-26555.
SHS thanks the Max Planck Society for support through the Max Planck Fellowship. This research is supported in part by the Excellence Cluster ORIGINS which is funded by the Deutsche Forschungsgemeinschaft (DFG, German Research Foundation) under Germany's Excellence Strategy -- EXC-2094 -- 390783311.
MM acknowledges support by the SNSF (Swiss National Science Foundation) through mobility grant P500PT\_203114.
SS has received funding from the European Union’s Horizon 2022 research and innovation programme under the Marie Skłodowska-Curie grant agreement No 101105167 — FASTIDIoUS.
This research is based in part on observations from the NASA/ESA Hubble Space Telescope program GO-15320, which is operated by the Space Telescope Science Institute. 
This research made use of \textsc{NumPy} \citep{oliphant2015,harris+2020} and \textsc{SciPy} \citep{jones+2001,virtanen+2020}. This research made use of \textsc{matplotlib}, a 2D graphics package used for \textsc{Python} \citep{hunter2007}.
\end{acknowledgements}

\bibliographystyle{aa}
\bibliography{tdcosmo_wgd2038_cosmo}

\begin{appendix}

\section{Results without $\kext$ or kinematics}\label{app:nokext_nokin}
We present the results of the $\tdist$ inference without including the external convergence ($\kext$) or stellar kinematics in Table~\ref{tab:ddt_nokext_nokin}.  We provide the constraints for the power-law and composite results, but not the combined constraints between the two, as removing the kinematics information changes the relative weighting of the models for {\sc lenstronomy}.  These results will be included in an upcoming hierarchical analysis of the entire TDCOSMO sample (TDCOSMO collaboration, in preparation), in which $\kext$ and kinematics will be incorporated for treatment at the population level.

\renewcommand*\arraystretch{1.5}
\begin{table}[h]
\caption{Time-delay distance constraints, without $\kext$ or kinematics included.\label{tab:ddt_nokext_nokin}}
\begin{minipage}{\linewidth}
\begin{tabular}{l|c}
\hline
\multirow{2}{*}{Model} &
$D_{\Delta t}$
\\
&
(Gpc)
\\
\hline
{\sc glee} power-law &
$1.47_{-0.32}^{+0.31}$\\
{\sc glee} composite &
$1.64_{-0.35}^{+0.35}$\\
\hline
{\sc lenstronomy} power-law &
$1.52_{-0.33}^{+0.34}$\\
{\sc lenstronomy} composite\footnote{The \textsc{lenstronomy} composite model is highly excluded by the kinematics data if $\lambda_{\rm int} = 1$ is assumed. Therefore, this value should not be used under this assumption. However, we provide this value for future usage with freely varying $\lambda_{\rm int}$.} &
$1.86_{-0.41}^{+0.41}$\\
\hline
{\sc glee}+{\sc lenstronomy} power-law &
$1.49_{-0.33}^{+0.32}$\\
{\sc glee}+{\sc lenstronomy} composite &
$1.74_{-0.39}^{+0.40}$\\
\hline
\end{tabular}
\\
\tablefoot{Reported values are medians, with errors corresponding to the 16th and 84th percentiles.  A flat prior in the range $0 \leq \tdist \leq 4~\mathrm{Gpc}$ is assumed.}
\end{minipage}
\end{table}
\renewcommand*\arraystretch{1.0}

\end{appendix}

\end{document}